\documentclass{article}

\usepackage{graphicx}        
\usepackage{multicol}        
\usepackage[bottom]{footmisc}

\usepackage{amsmath,amssymb,latexsym}
\usepackage{caption,subfigure}

\newcommand{\R}{\mathbb R}
\newcommand{\N}{\mathbb N}

\newcommand{\norm}[1]{\left\Vert#1\right\Vert}
\newcommand{\abs}[1]{\left\vert#1\right\vert}

\newcommand{\eps}{\varepsilon}

\begin{document}

\title{Fostering Consensus in Multidimensional Continuous Opinion Dynamics under Bounded Confidence}

\author{Jan Lorenz\\Universit{\"a}t Bremen, Bibliothekstr. 1, 28359 Bremen, Germany\\
\texttt{math@janlo.de}, \texttt{www.janlo.de}}
%
%
\date{First presented Aug 4, 2006 at Conference \\"Potentials of Complexity Science for Business, Governments, and the Media", Budapest}

\maketitle

\section{Introduction}
Consider a group of agents which is to find a common agreement
about some issues which can be regarded and communicated as real
numbers. Each agent has an opinion about each issue which he may
change when he gets aware of the opinions of others. This process
of changing opinions is a \emph{process of continuous opinion
dynamics}. Examples for discussing groups are parliaments, a commissions
of experts or citizens in a participation process. The opinion
issues in parliaments can be tax rates or items of the budget
plan, in commissions of experts predictions about macroeconomic
factors and for citizens the willingness to pay taxes or the
commitment to a constitution.

In many processes of opinion dynamics it is desirable that the agents reach \emph{consensus}, either for reaching a good approximation to the truth or for the reason, that reaching consensus is a good in itself (e.g. in the commitment to the constitution). Often, all relevant information about a societal issue has been collected and published but it is not reliable enough to bring a collective opinion or 'the truth' without opinion dynamics where agents judge, communicate and negotiate about the 'right' opinion. In the need of a collective decision it is the best for the group to achieve consensus because it does not need a decision by voting or other mechanisms with potential to conflict. In this study we make simple but reasonable assumption on humans in opinion dynamics. The models reproduce the formation of parties and interest groups and some other reasonable facts in real opinion dynamics. But there remain many reasonable free parameters of opinion dynamics, where we check a few with the aim to find structural conditions which might foster the achievement of consensus in the group. 

We define the models based on two facts from social psychology. First, people adjust their opinions towards the opinions of others. This may be for normative or for
informational reasons. So either because they feel conformational
pressure and want to assimilate or because they appreciate the
information of others to be relevant. Further on, people perceive
themselves as members of a subgroup, according to the theory of
self categorization. In our setting one feels as in a group with
the people who have similar opinions. We put these descriptions of peoples behavior regarding opinion dynamics into rules for agents behavior: Agents find new opinions as averages of opinions of others and they will do this only with respect to agents which lie within
their area of confidence. 

Repeated averaging and bounded
confidence lead to clustering dynamics. If the agents in our model have big enough areas of confidence they are able to find a consensus. If they are small they will
fail and form several clusters. Are their structural properties of the opinion dynamics environment that have a positive effect on the chances of finding a consensus?
Here, we will ask how structural properties of the opinion
dynamics process as the communication regime, the number of
opinion issues, their interdependence and the mode how agents form
their area of confidence affect the chances for consensus?

With the question about conditions for consensus we grab an old
research line of DeGroot \cite{DeGroot1974} and Lehrer and Wagner
\cite{Lehrer1981} about the problem how to aggregate opinions to a
rational consensus in science or society. They model aggregation
by averaging with powers of reputation matrices. The work in \cite{Lehrer1981} was in the flavor
of the social choice problem. In recent times
Hegselmann and Krause \cite{Hegselmann2002} grabbed on this with
the idea of bounded confidence and formulated a model (now
nonlinear) of opinion dynamics which can be seen as repeated
meetings of agents with bounded confidence. Independently, Weisbuch, Deffuant and others
\cite{Deffuant2000,Weisbuch2002} formulated a similar bounded
confidence model with random pairwise interaction, what we call
gossip communication. They came with the background of social
simulation, sociophysics and complexity science. 

In section \ref{sec2} we will outline and discuss the parameter
space and define the two opinion dynamic processes. Section
\ref{sec3} shows the basic dynamics which are universal in these
models: cluster formation in the time evolution and the
bifurcation of cluster patterns in the evolution of the bound of
confidence. We will set up on them in section \ref{sec4} where we
present and discuss the simulation results with a focus on the
consensus transition. We show e.g. that raising the number of opinion issues fosters consensus if the issues are under budget constraints, but diminishes consensus if they are not. We conclude by giving a colloquial summary
and pointing out further research directions.

\section{Continuous Opinion Dynamics und Bounded Confidence}\label{sec2}

Here, we define the basic models of \cite{Hegselmann2002} and
\cite{Weisbuch2002} such that they extend to more dimensional
opinions and to different areas of confidence. We briefly discuss
real world interpretations.

\paragraph{The agents}
Often, analytical results are either possible for very low numbers
of agents or in the limit for a large number of agents.
Complexity arises with finite but huge numbers of agents. The
fuzzy thing is that some macro level dynamics work, while at
critical points changes appear very sensitive due to specific
finite size effects. In the simulation studies we chose $n = 200$
because we regard this as applicable to a wide range of real
groups of agents. We also checked $n=50,500$ to ensure that the
results hold also in this range, which they do. This range of
group sizes coincides with the social brain hypotheses
\cite{HillDunbar2003} that humans can only hold about 150
relationships on average.

\begin{figure}
\centering
\includegraphics[width=0.9\textwidth]{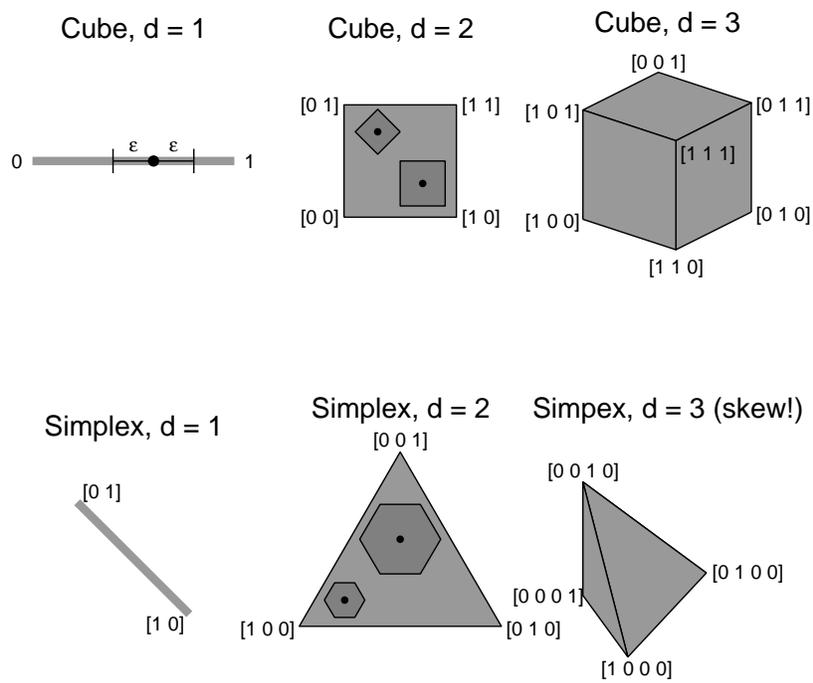}
\caption{$\Box$ and $\triangle$ opinion spaces with example areas
of confidence for $p=1,\infty$.}\label{figOpinionSpaces}
\end{figure}

\paragraph{The opinion space and the initial profile}
The opinion space is the set of all possible opinions an agent may
have. In continuous opinion dynamics about $d$ issues this is $\R^d$.
So, we call $x^i(t) \in \R^d$ the opinion of agent $i$ and $x(t)
\in (\R^d)^n$ the opinion profile at time $t \in \N$. The
evolution of an opinion profile is the \emph{process of continuous
opinion dynamics}. Dynamics depend heavily on the initial
opinion profile. If we model dynamics by repeated
averaging, then dynamics take place in the convex region spanned
by the initial opinion profile  $x(0)$, we call this the \emph{relevant opinion
space}. For $d=1$ this is always an interval. For higher $d$ there are
many shapes. In this study we will restrict us to $d = 1,2,3$ and
two shapes of the initial relevant opinion space: the \emph{cube}
$\Box^d:=[0,1]^d$ and the \emph{simplex} $\triangle^d :=
\{y\in\R^{d+1}_{\geq 0} \,|\, \sum_{i=1}^n y_i = 1 \}$ (see Fig. 
\ref{figOpinionSpaces}). Notice that $\triangle^d$ is a subset of
$\Box^{d+1}$. These two shapes stand for two different kinds of
multidimensional problems. The cube represents opinions about $d$
issues which can be changed independently. The simplex represents
opinions about issues where the magnitude of one can only be
changed by changing others in the other direction. The main
example is a budget plan with a fixed amount of money to allocate.
Further on, we restrict us to random initial opinions which are
equally distributed in the relevant opinion space. (It is not
trivial to produce an equal distribution on a simplex!
Normalization to sum-one of a $d+1$-dimensional cube would be
skewed. We produce it by taking a $d$-dimensional cube and
throwing away all opinions with sum bigger than 1. Then we compute
the missing least component for each opinion.)

\paragraph{The area of confidence}
The \emph{area of confidence} is a region in the opinion space
around an agent's opinion. He regards all opinions in this region
as relevant and all others as irrelevant. This region moves when
the agent changes his opinion. Formally, it is a compact and
convex subset of the opinion space including the origin.  The
origin is mapped to the opinion of the agent. In a one dimensional
opinion space the only relevant areas are intervals. In more
dimensions several areas seem appropriate. We restrict this study
to the unit balls of the $1$- and the $\infty$-norm (see Fig. 
\ref{figOpinionSpaces}) centered on the opinion and scaled by a
\emph{bound of confidence} $\eps > 0$. Thus, agents measure the
distance of opinions $x^1,x^2 \in \R^d$ as $\norm{x^1-x^2}_1 =
\sum_i \abs{x^1_i - x^2_i}$ or as $\norm{x^1-x^2}_\infty = \max_i
\abs{x^1_i - x^2_i}$ and judge their relevance by the threshold
$\eps$. We use these norms because they are close to how humans
may judge differences in opinion. Agents using the $1$-norm are
willing to compensate between the opinion issues. If the other
agent's opinion differs a lot in one issue this can be
\emph{compensated} by differing low in another issue. Agents using
the $\infty$-norm are \emph{noncompensators}. Their distance in
each opinion issue should be below $\eps$ to accept another's
opinion. For $d=1$ the area is always an interval. For the cube
and $d=2,(3)$ the $\infty$-ball is a square (cube); for the
$1$-ball it is a diamond (octahedron). The intersection of the
2-dimensional simplex and the 3-dimensional area of confidence is
a hexagon with edge length $\eps$ for the $\infty$-norm and with
edge length $\eps/2$ for the $1$-norm. The intersection of the
3-dimensional simplex and the 4-dimensional area of confidence is
an octahedron for the $\infty$-norm and a cuboctahedron for the
$1$-norm. Things get more fuzzy when going to more dimensions.

\paragraph{The communication regime}
The models of \cite{Hegselmann2002,Weisbuch2002} can both be
extended naturally to the different opinion spaces and the areas
of confidence outlined above. They differ in their communication
regime. In the model of Hegselmann and Krause
\cite{Hegselmann2002} each agent chooses his new opinion as the
arithmetic mean of all opinions in his area of confidence. All
agents do this at the same time. To do this, they need to know the
opinions of all agents. We call it communication by
\emph{repeated meetings}. In the basic model of Deffuant, Weisbuch
and others \cite{Weisbuch2002} two agents were chosen at random.
They compromise in the middle if their opinions lie in the area of
confidence of each other. We call this communication regime
\emph{gossip}. 

Now we are ready for the mathematical definition of the two 
processes of continuous opinion dynamics.

Given an initial profile $x(0) \in \R^{n}$, a bound of confidence
$\eps > 0$ and a norm parameter $p \in \{1,\infty\}$ we define the
\emph{repeated meeting process} $(x(t))_{t\in\N}$ recursively
through
\begin{equation}
x(t+1) = A(x(t),\eps) x(t),
\end{equation}
with $A(x,\eps)$ being the \emph{confidence matrix} defined
\[
 a_{ij}(x,\eps) :=
 \left\{ \begin{array}{cl}
   \frac{1}{\#I(i,x)} \quad & \textrm{if } j\in I(i,x)   \\
   0 & \textrm{otherwise,}
\end{array} \right. \\
\]
with $I(i,x) := \{j \,|\, \norm{x^i - x^j}_p \leq \eps \}$.
("$\#$" stands for the number of elements.)

We define the \emph{gossip process} as the random process
$(x(t))_{t\in\N}$ that chooses in each time step $t\in\N$ two
random agents $i,j$ which perform the action
\begin{eqnarray*}
x^i(t+1) &=& \left\{
\begin{array}{ll}
    x^i(t) + \frac{1}{2}(x^j(t)-x^i(t)) & \hbox{if $\norm{x^i(t)-x^j(t)}_p\leq\eps$} \\
    x^i(t) & \hbox{otherwise.}
\end{array}
\right.\\
\end{eqnarray*}
The same for $x^j(t+1)$ with $i$ and $j$ interchanged.

\begin{figure}
\centering
\includegraphics[width=0.3\textwidth]{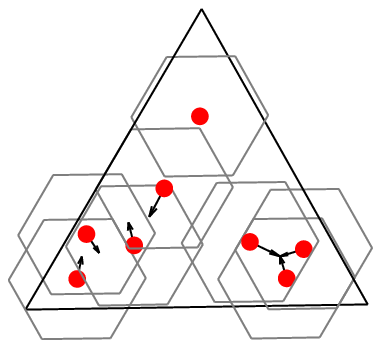}\qquad\qquad
\includegraphics[width=0.3\textwidth]{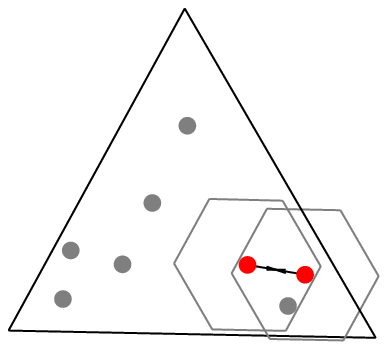}
\caption{Examples of one step in meeting (left hand) and gossip (right hand) dynamics in the opinion space $\triangle^2$.}\label{figExample2}
\end{figure}

Figure \ref{figExample2} demonstrates one time step in each process. 

\bigskip

\section{General Dynamics}\label{sec3}

\paragraph{Clustering dynamics in the time evolution}\label{subsec31} Every
gossip and meeting process converges to a fixed configuration of
opinion clusters \cite{Lorenz2005,Moreau2005}. We call this fixed
configuration the \emph{stabilized profile}. A general dynamic
is that opinion regions with high agent density attract agents
from around. This attraction comes due to a higher probability to
meet an agent in this region in gossip communication and due to
the fact that the barycenter of opinions in an area of confidence
is often close to a high density region.

\begin{figure}[htbp]
  \centering
  \subfigure[{Example processes for gossip and meeting communication in the interval $[0,1]$
  demonstrating the time evolution to a stabilized
  profile. Notice one outlier for gossip and the meta-stable state in meetings.}]{\label{subfig1}\includegraphics[width=0.49\textwidth]{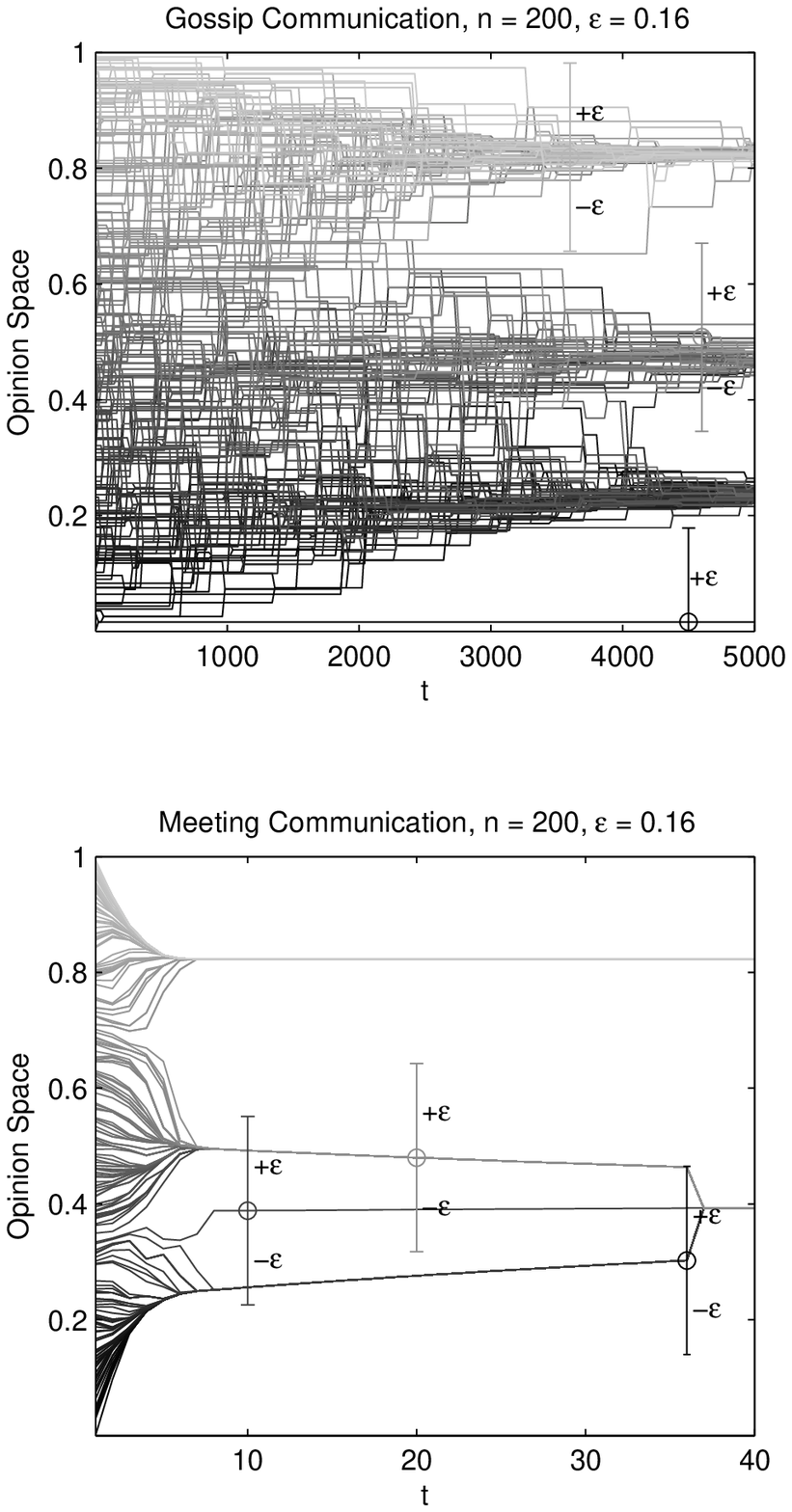}}
  \, \subfigure[{Reverse bifurcation diagrams of characteristic
  states of the stabilized profile in the $\eps$-evolution. Diagrams derived by interactive Markov chains.
  Black is a high number of agents, gray a low number of agents.}]{\label{subfig2}\includegraphics[width=0.48\textwidth]{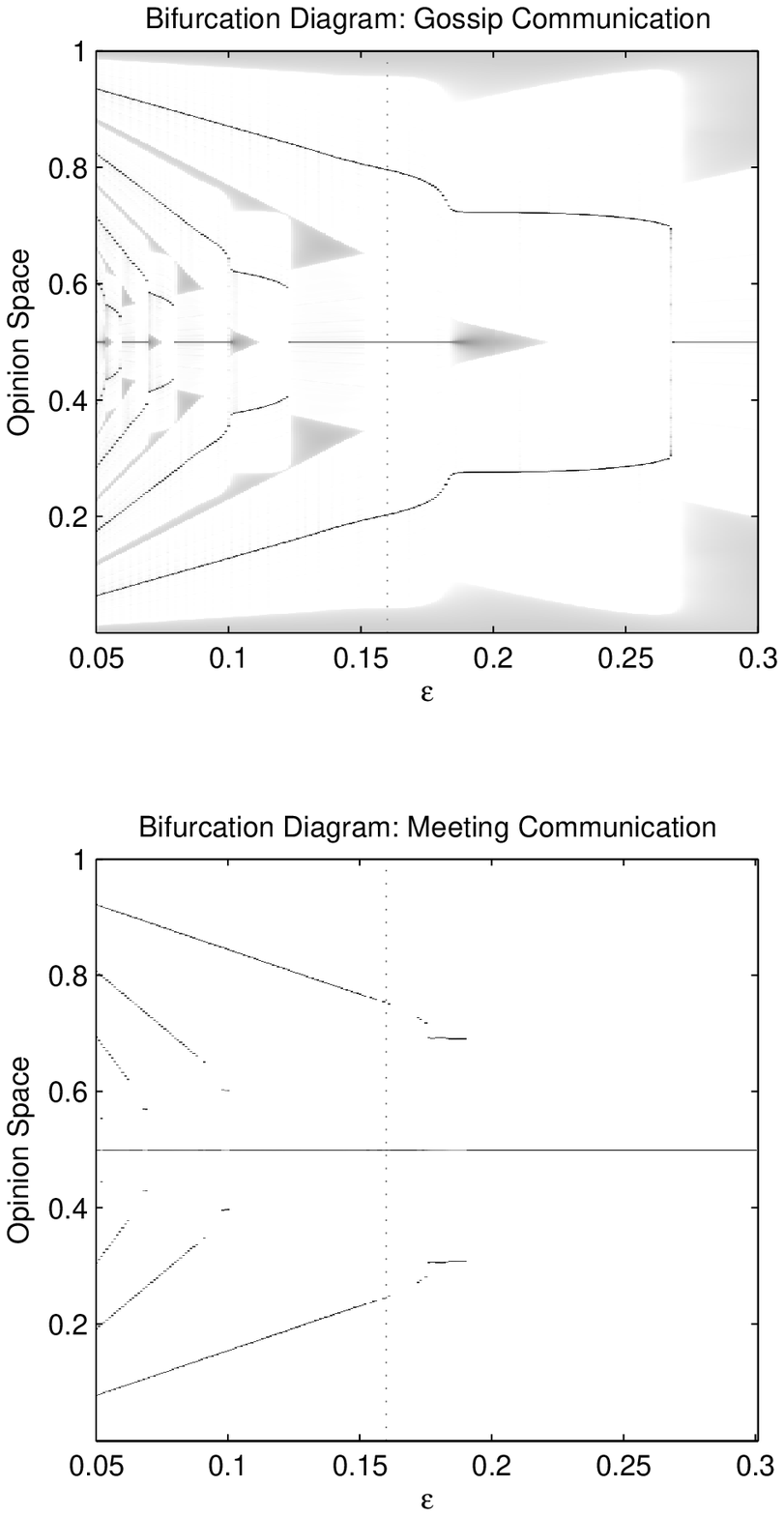}}
  \, \label{figRunBif}
  \caption{Demonstration of general dynamical properties.}
\end{figure}

\begin{figure}
  \includegraphics[width=\textwidth]{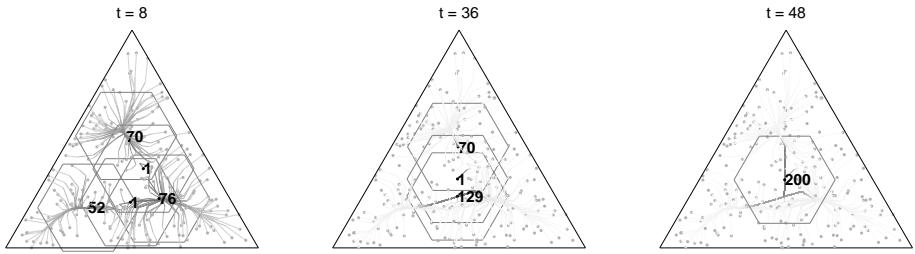}
  \caption{Example for meeting communication in
  $\triangle^2$ for interesting time steps. Notice the successive
  joining of intermediate clusters.}\label{figExSimp}
\end{figure}

 If we consider an initial
profile with uniformly distributed opinions on a certain relevant
opinion space ($\Box$ or $\triangle$) than the density
distribution of opinions evolves over time as follows. (The
following dynamic description can be traced in Fig. 
\ref{subfig1} for $\Box^1$ and for $\triangle^2$ in Fig. 
\ref{figExSimp}.) Agents at the border of the relevant opinion space
move closer to the center because opinions in their area of
confidence are not equally distributed. Density in the center
changes only due to random fluctuations in the initial conditions. So
the relevant opinion space contracts but holds mainly the same
shape but with a higher agent density at the border. If a more
dimensional opinion space had some vertices (as $\Box$ and
$\triangle$ have) the density in the evolving high density regions
is even higher at the vertices due to opinions coming from more
sides.

These high density regions at the vertices of the relevant opinion
space attract agents from the center and may get disconnected from
the center and from the other vertices at some time, due to
absorbtion of the connecting agents, and form a cluster. The
dynamics goes on similar in the remaining cloud of connected
opinions.

If some of these high density regions lie as close to each other
that a small group of agents holds contact to both, then it may
happen that they attract the agents in these high density regions
and both join to form a bigger cluster. This may also happen to
more clusters at the same time or with some delays (see Fig. 
\ref{figExSimp} for an example). The fuzzy thing in more dimensions
is that this contracting process happens on all face levels (e.g.
faces and edges) of the shape of the opinion space on overlapping
time scales. Further on, some clustering in the center may also
occur due to slow deviations of uniformity. The time when some
high density regions have formed but have not completely
disconnected from the rest is thus the critical time phase. In
more than one dimension it is unpredictable which of the
intermediate clusters joins with which others. Changes may happen
due to very low fluctuations in the initial profile or the
communication order.

$\triangle^d$ has $d+1$ vertices and thus the same number of intermediate high
density regions. The number of possible final cluster
configurations that may evolve by disconnecting or joining of these high density regions is the
same as the number of partitions of $\{1,\dots,d+1\}$ into pairwise disjoint subsets, which is the
\emph{Bell number} $B_{d+1}$. This shows the combinatorial
explosion of different possible outcomes: $B_2 = 2, B_3 = 5, B_4 =
15, B_5 = 52, B_6 = 203, B_7 = 877, B_8 = 4140$.

\paragraph{Bifurcation dynamics in the evolution of the bound of confidence}
For each value of $\eps$ there is a certain \emph{characteristic
stabilized profile} under the assumption of a uniformly
distributed initial profile. The number, the size and the location
of opinion clusters in this stabilized profile are of interest. In
Fig. \ref{subfig2} we see the reverse bifurcation diagrams for
the attractive states of the meeting and the gossip process in
$\Box^1=[0,1]$ as relevant opinion space. These diagrams have been
computed with interactive Markov chain that govern the evolution
of the distribution of an idealized infinite population to a huge
number of opinion classes in the opinion space (for details see
\cite{Lorenz2005,Lorenz2006} and \cite{Ben-Naim2003} for the
inspiring differential equation approach). Such bifurcation diagrams should exist for moredimensional
opinion spaces, too. A stabilized profile with 200 agents can
be significantly blurred by low fluctuations in the initial
profile and thus does not behave as the bifurcation diagram predicts. But as simulation shows, a bifurcation diagram with
attractive states and certain discontinuous changes when
manipulating $\eps$ seems to underlie opinion dynamics under
bounded confidence.

It is easy to accept that $\eps = 1$ leads to a
central consensus, while $\eps\to 0 $ leads to full plurality
where no opinion dynamic happens. The behavior in between can be
understood as \emph{bifurcations} of the consensual central
cluster into other configurations of clusters. In the gossip and
the meeting process the main effect when going down with $\eps$ is
that the central cluster bifurcates at certain values of $\eps$
into two equally sized major clusters left and right which drift
outwards when lowering $\eps$ further. The central cluster
vanishes (nearly) completely to get reborn and grow again until it
bifurcates again. We call the interval between two bifurcation
points an \emph{$\eps$-phase} for a characteristic stabilized
profile. The length of the $\eps$-phases scales with $\eps$, so
for lower $\eps$ the phases get shorter. This fact is the basis of
the $1/2\eps$-rule (see \cite{Weisbuch2002}) which determines the
number of major clusters under gossip communication.

Besides the common behavior the gossip and the meeting process
differ. Under gossip communication there are minor clusters at the
extremes, a nucleation of minor clusters between the central and
the first off-central clusters and minor clusters between two
major off-central clusters. These minor clusters occur as a few
outliers in agent based example processes, too. Meeting
communication shows no minor clusters but the surprising phenomena
of consensus striking back after bifurcation. Convergence in this
phase takes very long (see \cite{Lorenz2006}). The long
convergence times to central consensus occurs also in front of
each bifurcation of the central cluster. E.g. for $\eps=0.2$ we
reach a meta-stable state of two off-central clusters and a small
central cluster which attracts them very slowly to a consensus.
The slow convergence due to meta-stable states close to
bifurcation points occurs also in example processes.

In this study we focus on fostering consensus. So the most
interesting point for us is the value of $\eps$ where the big
central cluster bifurcates into two major clusters. This is the
phase transition from polarization to consensus. We call this the
\emph{majority consensus transition}. Only 'majority' not total
because of the extremal minor clusters in gossip communication. We
call this point (in the style of \cite{Hegselmann2004a}) the
\emph{majority consensus brink}.

\section{Simulation Results}\label{sec4}

\paragraph{Simulation setup} Our simulation setup deals with
initial profiles of random and equally distributed opinions with
200 agents. We run processes for the 24 settings of the opinion
spaces $\Box,\triangle$ with dimensions $d=1,2,3$, the areas of
confidence for $p=1,\infty$ and the communication regimes meeting
and gossip. For each of these settings we took a big enough range
of $\eps$-values in steps of $0.01$ so that we are sure that the
majority consensus transition happens within this range. For each
of this 24 settings and each value of the respective $\eps$-range
we run 250 simulation runs and collect the stabilized profiles for
our final statistical analysis. We checked 50 and 500 agents with
lower numbers of runs and verified that the results hold analog
qualitatively and to a large extend quantitatively.

For each collection of stabilized profiles for a given point in
the
$\{\Box,\triangle\}$-$\{d=1,2,3\}$-$\{p=1,\infty\}$-$\{$meeting/gossip$\}$-$\eps$-parameter
space, we have to measure the degree of consensus. In earlier
studies the most used measure was the average number of clusters.
This is inappropriate because of the minor clusters at the
extremes under gossip communication. We use the \emph{average size
of the biggest cluster}. If it is 200 we are for sure above the
majority consensus brink. If it is slightly below this can have
two reasons according to what we know from section \ref{sec3}.
First, some runs reach consensus, while some others polarize, or
second, there is a big central cluster but also an amount of
agents in minority clusters at the extremes. The second happens
mostly for gossip communication.

\begin{figure}[htbp]
\centering
\includegraphics[width=0.9\textwidth]{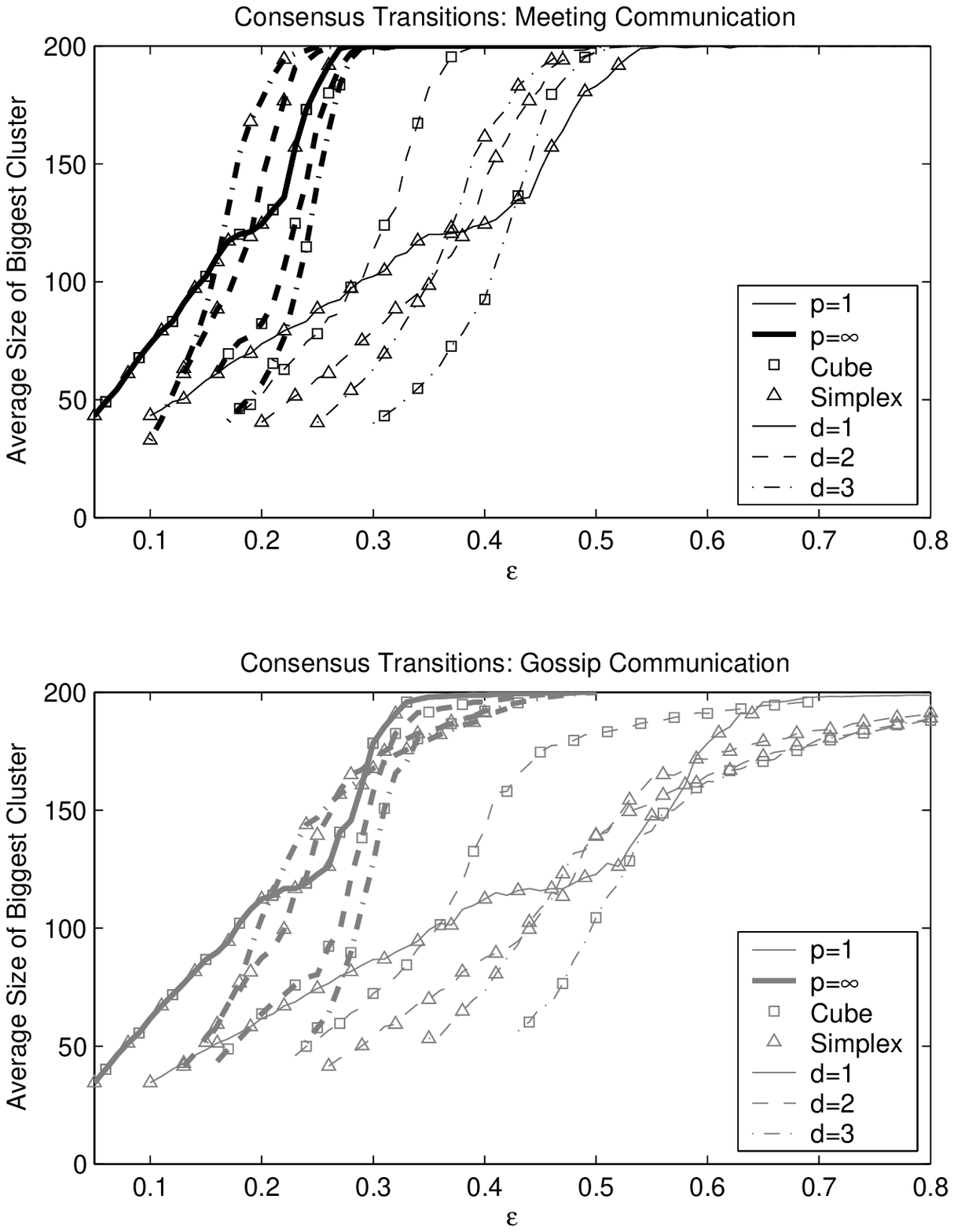}
  \caption{The average size of the biggest cluster for $\Box,\triangle$ (marker), $d=1,2,3$ (line style), $p=1,\infty$ (line width) and communication regime (black, gray).}\label{subfig4}
\end{figure}

\begin{figure}[htbp]
  \centering
  \subfigure[{The lines of Fig. \ref{subfig4} with $\eps$-axis
  for lines of meeting communication scaled to $80\%$.}]{\label{subfig5}\includegraphics[width=0.9\textwidth]{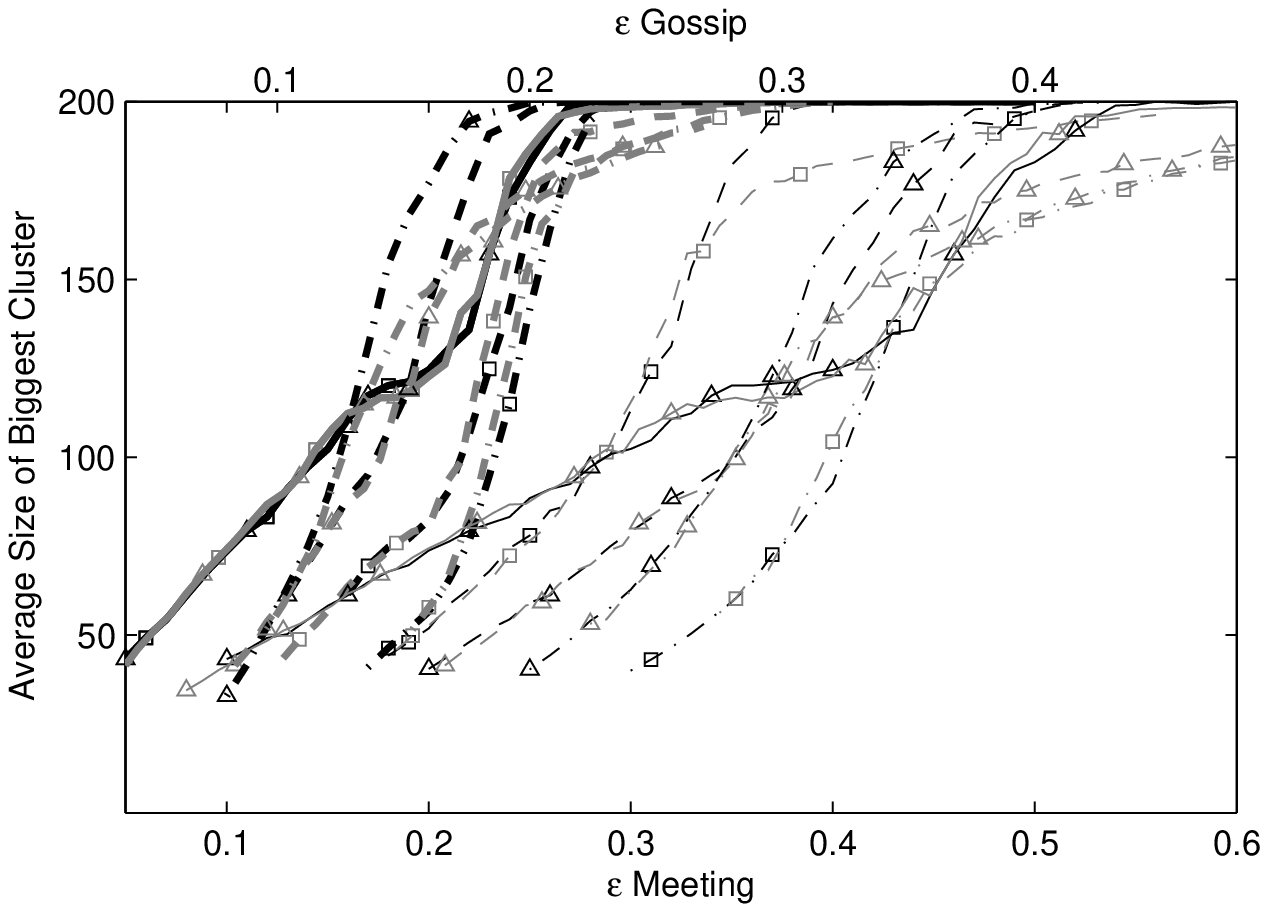}}
  \, \subfigure[{The lines of Fig. \ref{subfig4} with $eps$-axis
  for lines of $p=1$ scaled to volume equality with $p=\infty$.}]{\label{subfig6}\includegraphics[width=0.9\textwidth]{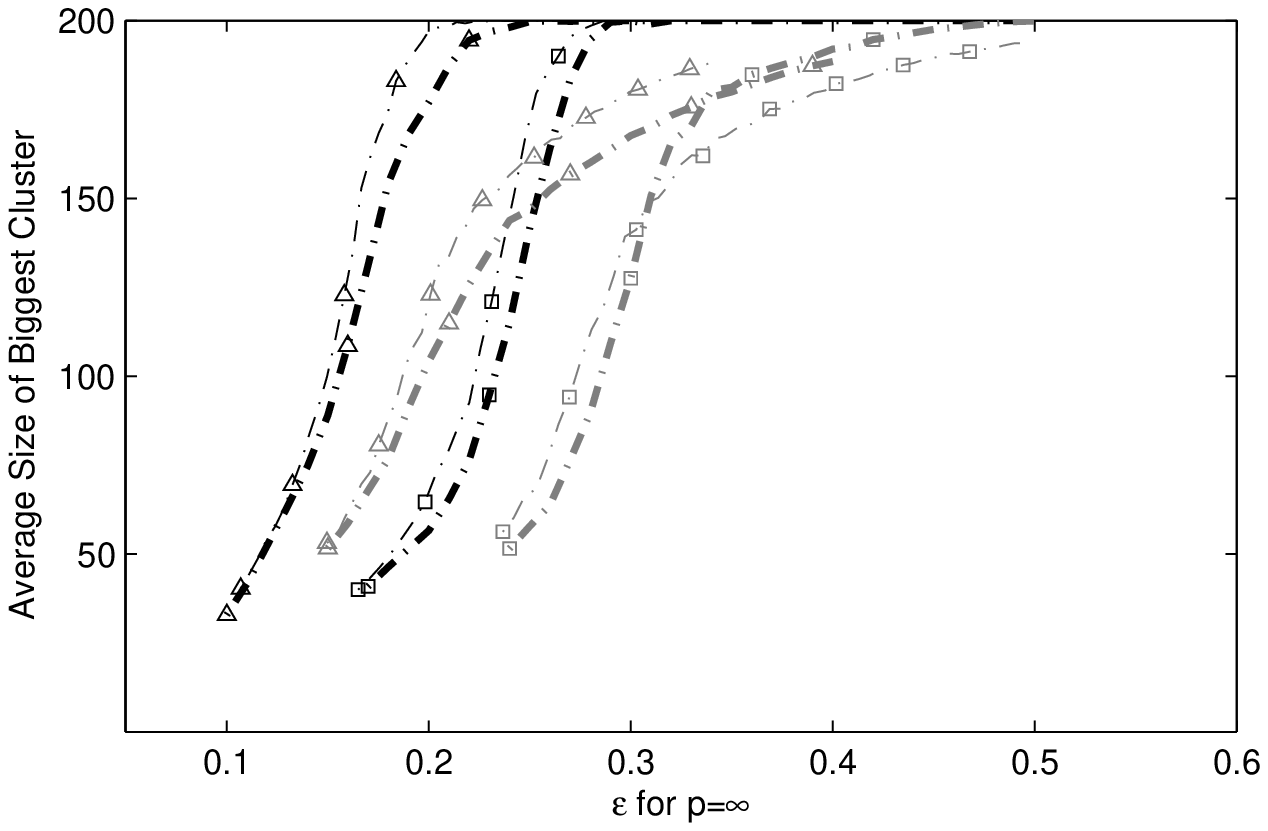}}
  \caption{Further simulation results for the average size of the biggest cluster.}
\end{figure}

Figure \ref{subfig4} shows the average size of the biggest cluster
with respect to $\eps$ for all 24 parameter setting. We derive
qualitative statements about \emph{fostering consensus} from that.
With fostering consensus we mean that the transition to a majority
consensus appears for lower values of $\eps$.

\paragraph{The impact of the communication regime (meetings vs.
gossip)} Communication in repeated meetings is fostering consensus
in comparison to gossip. But surprisingly Fig. \ref{subfig5}
gives strong evidence about the universal scale that a group of
agents in meeting communication needs only $0.8\eps$ to reach the
same average size of the biggest cluster as the same group under
gossip communication with $\eps$. This holds for all our parameter
settings. Only for very high sizes of the biggest cluster meeting
communication gets even better, probably due to more minor
clusters in gossip communication.

\paragraph{The impact of the number of opinion issues $d$}
What happens if we raise the number of issues? The answer is: It
depends on the shape of the initial relevant opinion space. In a
simplex, raising the number of issues fosters consensus. In a cube
raising the number of issues dilutes consensus. Numerical values
for fostering with meeting communication in a simplex and
$p=\infty$: the biggest cluster contains at least $80\%$ of the
agents in $80\%$ of the runs for $\eps > 0.25$ with $d=1$, $\eps >
0.23$ with $d=2$ and , $\eps > 0.20$ with $d=3$. One drawback is
that under gossip communication we produce more and bigger
extremal minor clusters in a simplex when raising $d$, one in each
vertex. Thus, for fostering a complete consensus without outliers raising
dimensionality under gossip dynamics is not good.

\paragraph{The impact of the shape of the relevant opinion space ($\triangle$ vs. $\Box$)}
What fosters consensus better: an opinion space of three
independent issues ($\Box^3$) or four issues under fixed budget
constraints ($\triangle^3$)? Colloquial: Is it good to add a budget dimension. The simplex is better for all $p$ and
all communication regimes. But this does not hold for $d=2$, where
the square is better under $p=1$ but the simplex is better under
$p=\infty$. Both shapes are trivially equal for $d=1$. We
conjecture that the simplex is getting better in higher
dimensions. Another question of similar type is: Does it foster
consensus to break a problem of three independent issues
($\Box^3$) down to a problem of three issues under budget
constraints ($\triangle^2$)? The answer is yes. It holds also for
breaking down from $\Box^2$ to $\triangle^1$ under $p=\infty$, but
it is the other way round for $p=1$.

\paragraph{The impact of compensating vs. noncompensating ($p =
1,\infty$)} Imagine you appeal to your noncompensating
($p=\infty$) agents 'compensate: switch to $p=1$'. This would
imply that they should not tolerate distances of $\eps$ in each
issue but only in the sum of all distances. Of course this will
not foster consensus because their area of confidence is then only
a smaller subset of their former. Perhaps you can appeal, that
they should compensate in the way such that they should allow
longer distances then $\eps$ in one issue in the magnitude as the
other distances are short. This would lead to maximal distances of
$d\eps$ in one issue and perhaps the agent find this two much to
tolerate. The 'mathematically correct' switching from
noncompensating to compensating is to scale $\eps$ to that
magnitude that the $d$-dimensional volumes of the areas of
confidence would be equal. We did this for $d=3$ in Fig. 
\ref{subfig6}. The scale for $\Box^3$ is $\sqrt[3]{6} \approx
1.82$ and for $\triangle^3$ it is $\sqrt[3]{64/5} \approx 2.34$.
This 'normalization' leads to the result that switching to
compensating fosters consensus a little bit. Probably this result
holds only in this configuration of the relevant opinion space and
the area of confidence, there might be negative configurations.

\section{Summary and Outlook}\label{sec5}

A colloquial summary: If we want to foster consensus and believe
that agents adjust there opinions by building averages of other's
opinions but have bounded confidence, then we should manipulate
the opinion formation process in the following way (if possible):
\begin{itemize}
\item Install meetings (or publications) where everyone hears all
opinions and do not rely only on gossip.

\item Bring more issues in but put them under budget constraints.

\item Release guidelines about compensation in the judgements of
different issues.
\end{itemize}
Of course, our simple model neglects several properties of real
opinion dynamics, e.g. rules about voting decisions, underlying
social networks, heterogeneity of agent's confidence, long run
ideologies or strategies and inflow of new information. All this
are tasks for further analytical and experimental work. An
unanswered question is also the reason for the universal $80\%$
scale for meeting communication compared to gossip.

But we believe that under more realistic extensions there will be
influence of the underlying bifurcation diagram and that critical
consensus transitions will exist. Thus, it is worth to observe and
design the structural properties of opinion dynamic processes, if
one aims to foster consensus.



\end{document}